\documentclass[prd,aps,eqsecnum,amsmath,amssymb,floatfix,nofootinbib,preprint,preprintnumbers,tightenlines]{revtex4}
\usepackage{amsmath,amssymb, bm}
\usepackage{dsfont}
\usepackage{epsfig}
\usepackage{graphicx}
\usepackage{slashed}
\usepackage{color}
\usepackage{subfigure}

\usepackage[colorlinks,citecolor=blue,urlcolor=blue,linkcolor=blue]{hyperref}

\newcommand{\be}{\begin{equation}}
\newcommand{\ee}{\end{equation}}

\newcommand{\beq}{\be}
\newcommand{\eeq}{\ee}
\newcommand{\bea}{\begin{eqnarray}}
\newcommand{\eea}{\end{eqnarray}}
\newcommand{\ba}{\begin{align}}
\newcommand{\ea}{\end{align}}
\newcommand{\bfig}{\begin{figure}}
\newcommand{\efig}{\end{figure}}

\newcommand{\wh}{\widehat}
\newcommand{\nn}{\nonumber}

\begin{document}

\font\rm=cmr12
\font\bf=cmbx12
\font\it=cmti12
\rm
\thispagestyle{empty}

\title{Power corrections to the modified QCD perturbative  series based on conformal mapping  of the Borel plane}

\vspace{3ex}
\author{Irinel Caprini}

\affiliation{Horia Hulubei National Institute for Physics and Nuclear Engineering, P.O.B. MG-6, 077125 Bucharest-Magurele,
 Romania\vspace{1cm}}

\begin{abstract}
\vskip0.5cm
 
Modifications of the QCD perturbative expansions  by the subtraction of the dominant infrared renormalon have been proposed recently  as  attempts to solve the long-standing discrepancy between fixed-order and contour-improved  perturbation theory for the hadronic $\tau$ decays. In this approach, the modified  perturbative series is supplemented by a modified gluon condensate in the operator product expansion of the  Adler function. Motivated by these works, we revisited recently a formulation of the QCD perturbation theory, proposed some time ago, which takes into account the renormalons by means of the conformal mapping of the Borel plane. One expects that the modified perturbative series obtained in this framework should be accompanied by modified  power-suppressed nonperturbative corrections. However, in the previous studies  the focus was on the perturbative series and the question of the possible power corrections has not been considered. In the present paper, we investigate for the first time this problem. Using techniques  from the mathematical resurgence theory, we derive the expression of the dominant power correction to the perturbative series obtained by conformal mapping of the Borel plane, and discuss its  implications for phenomenological applications. 
 
\end{abstract}

\maketitle 
\vspace{0.2cm} 
\section{Introduction}
 
 The data on the inclusive hadronic decay width of the $\tau$ lepton allow the measurement of the
strong coupling $\alpha_s(\mu^2)$  at the relatively low scale $\mu = m_\tau$. By analyticity, the $\tau$ hadronic width is expressed as a weighted integral of the Adler function $D(s)$ along the circle $|s|=m_\tau^2$ in the complex $s$ plane, where one can use perturbation theory and the operator product expansion (OPE) \cite{Braaten:1991qm}. The perturbative part can be   expanded in powers of $\alpha_s(m_\tau^2)$ \cite{Braaten:1991qm}, a formulation known as fixed-order perturbation  theory (FOPT), or the powers of the running coupling  $\alpha_s(-s)$ can be integrated along the circle  \cite{LeDiberder:1992zhd},  an approach known as contour-improved perturbation theory (CIPT). 
 
Contrary to expectations based on renormalization-group invariance,  there is a significant numerical difference between the results of the two methods  \cite{ParticleDataGroup:2022pth}.  This  discrepancy, which sets the most important limitation on the precision of $\alpha_s$ extracted from hadronic $\tau$ decay, has been discussed in many papers. In particular, in \cite{Beneke:2008ad} it was noticed that the  difference does not decrease when more terms are included in the perturbative series, the reason being the fact that the QCD perturbative expansions are divergent series.  More recently, the crucial role of the renormalons, which encode the large-order behavior of the perturbative series \cite{Beneke:1998ui}, has been emphasized in several papers. Thus, in   \cite{Benitez-Rathgeb:2022yqb, Benitez-Rathgeb:2022hfj, Beneke:2023wkq}, the leading infrared renormalon divergence related to the gluon condensate was subtracted from the  perturbative series of the Adler function. By this modification,  the difference between the fixed-order and the contour-improved perturbative expansions was considerably reduced. On the other hand, the  modified perturbative series  is accompanied in this approach by a modified gluon condensate in the OPE.

Stimulated by these works, in  the recent paper \cite{Caprini:2023tfa} we brought into attention a different modification of the QCD perturbation theory, which  also takes into account the renormalons. The new expansions,  proposed for the first time in \cite{Caprini:1998wg} and investigated further in \cite{Caprini:2000js, Caprini:2001mn, Caprini:2009vf, Caprini:2011ya, Abbas:2013usa, Caprini:2019kwp, Caprini:2020lff, Caprini:2021wvf}, are obtained by  expanding the Borel transform of the Adler function in powers of a new variable, which performs the  conformal mapping of the Borel plane, with cuts along the real axis due to renormalons, onto the unit disk of a new complex plane. The original function is then recovered from the Borel transform by Laplace-Borel integral with principal value (PV) prescription. In  \cite{Caprini:2023tfa} we revisited some of the properties of these expansions, and compared the method of conformal mapping with the modified series proposed in  \cite{Benitez-Rathgeb:2022yqb, Benitez-Rathgeb:2022hfj, Beneke:2023wkq}. In particular, since the expansions based on conformal mapping have a tamed behavior at large orders, they lead to a reduced discrepancy between the fixed-order and contour-improved formulations.
  
 In the previous studies of the method,  the primary focus was  on the perturbative series, and the modification of the power corrections was not mentioned as a dedicated aim.  But, as  remarked in  \cite{Benitez-Rathgeb:2022yqb},  one may expect that this approach represents effectively a realization of a renormalon-free OPE scheme.  In the present paper we investigate  this problem, and derive for the first time the specific form of the power corrections to the series based on conformal mapping of the Borel plane. 
 
 Since the perturbative series in QCD is not Borel summable due to the infrared renormalons, which  are located on the positive axis of the Borel plane, we resort to  the mathematical theory of Borel nonsummable series, the so-called resurgence theory. In this approach, the sought function is expected to ``resurge`` by supplementing the perturbative series  with terms singular at the expansion point, which cannot be seen perturbatively. These terms, denoted generically as ``nonperturbative'', can be organized sometimes as a ``transseries'', i.e., a sequence of power series, the expansion parameter of each series being exponentially suppressed with respect to that of the previous series. Remarkably, insights on the form of the additional terms can be obtained from perturbation theory itself, more exactly from the large-order behavior of the perturbative asymptotic expansion.  
 
 Several mathematical approaches to resurgence and transseries have been proposed (see \cite{Ecalle1993, BerryHowls, Howls, Dorigoni:2014hea, Costin1995, Costin_Duke, Costin_Book,  Aniceto:2018bis} and references therein), and there are some applications of these mathematical methods to QCD.  One such application is developed in a series of papers \cite{Ayala:2019uaw, Ayala:2019hkn, Ayala:2020pxq}, where the truncated perturbation series of some QCD correlators have been supplemented by nonperturbative terms related to  the first infrared renormalon, in the spirit of the hyperasymptotic formalism from \cite{BerryHowls, Howls}. Other application is the resurgent representation of the Adler function in the large-$\beta_0$ approximation,  derived in \cite{Maiezza:2021mry, Caprini:2023kpw}, based on the previous works \cite{Maiezza:2019dht, Bersini:2019axn}, where is was shown how the renormalization-group equation satisfied by the Adler function ensures the applicability to this function of the resurgence approach developed in \cite{Costin1995, Costin_Duke, Costin_Book}.    
 
  On the other hand, in phenomenological analyses,  the QCD  perturbative series has been supplemented since a long time  by power-suppressed corrections, which can be viewed as terms singular at the origin of the coupling plane. They have been proposed for the first time in  \cite{Shifman:1978bx} by extending the validity of the OPE to the nonperturbative regime, and are expressed in terms of nonzero quark and gluon vacuum matrix elements (condensates). However, one must keep in mind that the validity of the OPE is proven rigorously only within  perturbation theory,  its extension to the nonperturbative regime being a conjecture. One must recall also that the power corrections depend of the formulation of the perturbative series. For instance, the redefinition of the perturbative series proposed in \cite{Benitez-Rathgeb:2022yqb, Benitez-Rathgeb:2022hfj,  Beneke:2023wkq},  is accompanied by a simultaneous redefinition  of the gluon condensate. 

 Our aim in this paper is to derive the power corrections for the modified QCD perturbative series based on the conformal mapping of the Borel plane. To achieve this goal, we resort to the mathematical theory of resurgence developed in \cite{Costin1995, Costin_Duke, Costin_Book} for  the solutions of differential equations. In this framework, the nonperturbative corrections are obtained from the discontinuity of the Borel transform  across the lines of singularities (Stokes lines) in the Borel plane.  As shown in \cite{Maiezza:2019dht, Bersini:2019axn}, the applicability of the method to the QCD Adler function is ensured by the renormalization-group equation satisfied by this function.  Actually, as mentioned above, the algorithm developed in \cite{Costin1995, Costin_Duke, Costin_Book} was applied in  \cite{Maiezza:2021mry, Caprini:2023kpw} for deriving the resurgent representation of the Adler function in the large-$\beta_0$ approximation.

The outline of the paper is as follows: in the next section we briefly review the standard perturbative expansion of the QCD Adler function and the modified expansion based on the conformal mapping of the Borel plane. In Sec. \ref{sec:OPE} we illustrate the application of the algorithm proposed in \cite{Costin1995,Costin_Duke, Costin_Book} to the standard perturbative expansion, assuming for simplicity that the Borel transform consists from a single branch point. In Sec.  \ref{sec:OPE1} we apply the same algorithm to the series based on conformal mapping and derive the form of the power corrections in this scheme. In Sec. \ref{sec:phen} we briefly discuss some numerical and phenomenological implications of our results.  Finally, Sec. \ref{sec:conc} contains the summary of the work and our conclusions. 

\section{Perturbative expansions of the Adler function}\label{sec:Adler}
 We  consider the reduced Adler function \cite{Beneke:2008ad}
\beq\label{eq:D}
\widehat{D}(s) \equiv 4 \pi^2 D(s) -1,
\eeq
where $D(s)=-s \,d\Pi(s)/ds$ is the logarithmic derivative of the invariant amplitude $\Pi(s)$ of the two-current correlation tensor.  From general principles of field theory, it is known that $\wh D(s)$ is an analytic function of real type (i.e., it satisfies the  Schwarz reflection property $\wh D(s^*)=\wh D^*(s)$) in the complex $s$ plane cut along the timelike axis for $s\ge 4 m_\pi^2$. 

 In QCD perturbation theory,  $\wh D(s)$ is expanded as  
 \be\label{eq:hatD1}
\widehat{D}(s) =\sum\limits_{n\ge 1} c_{n,1}\, [a_s(-s)]^n,
\ee
in powers of the strong running coupling $a_s(-s)\equiv \alpha_s(-s)/\pi$,  which satisfies the renormalization-group equation 
 \be\label{eq:rge}
 -\mu\frac {d a_s}
{d\mu}\equiv \beta(a_s)=\sum_{n\ge 1}
\beta_n a_s^{n+1} \ee
 in a certain renormalization scheme. The  coefficients  $c_{n,1}$ are real, since $\widehat{D}(s)$ satisfies Schwarz reflection.  They are known for $n\le 4$  in the $\overline{\mbox{MS}}$ scheme, cf. \cite{Baikov:2008jh} and references therein:
 \be\label{eq:cn1}
 c_{n,1}=1,\quad c_{2,1}=1.64;\quad c_{3,1}=6.37, \quad c_{4,1}=49.07.
\ee
At high orders, the coefficients increase factorially,  $c_{n,1}\sim  n !$  \cite{Beneke:1998ui}. Therefore, the series (\ref{eq:hatD1})  has zero radius of convergence and can be interpreted only as an asymptotic  expansion to $\widehat{D}(s)$ for $a_s\to 0$. This indicates the fact that the Adler function, viewed as a function of the strong coupling $a_s$, is singular at the origin $a_s=0$ of the coupling plane. 
 
In some cases, the expanded functions can be recovered  from their divergent expansions through Borel summation. The Borel transform of the Adler function is defined by the power series
\be\label{eq:B}
 B_{\widehat D}(u)= \sum_{n=0}^\infty  b_n\, u^n,
\ee
where the coefficients $b_n$ are related to the perturbative coefficients $c_{n,1}$ by 
\be\label{eq:bn}
 b_n= \frac{c_{n+1,1}}{\beta_0^n \,n!}..
\ee
Here we used the standard notation $\beta_0=\beta_1/2$, and in our convention $\beta_0=9/4$.

The large-order increase of the coefficients of the series  (\ref{eq:hatD1}) is encoded  in the singularities of the Borel transform in the complex $u$ plane. In the present case, it is known that $B_{\wh D}(u)$ has singularities at integer values of $u$ on the semiaxes $u\ge 2$ (infrared renormalons), and $u\le -1$ (ultraviolet renormalons)  \cite{Beneke:1998ui}.  In a particular limit of perturbative QCD, known as large-$\beta_0$ approximation 
\cite{Beneke:1992ch, Broadhurst:1992si, Beneke:1994qe},  the singularities are poles, but in full QCD they are branch points.

From the definition (\ref{eq:B}), it follows that the function $\wh D(s)$  can be obtained formally from the Borel transform  $B_{\widehat D}(u)$ by the Laplace-Borel integral  
\be\label{eq:Laplace}
\wh D(s)=\frac{1}{\beta_0} \,\int\limits_0^\infty  
\exp{\left(\frac{-u}{\beta_0 a_s(-s)}\right)} \,  B_{\wh D}(u)\, d u.
\ee
Actually, due to the singularities of $ B_{\wh D}(u)$ for $u\ge 2$, the  integral (\ref{eq:Laplace}) is not defined and requires a regularization. As shown in \cite{Caprini:1999ma}, the PV  prescription, where the integral (\ref{eq:Laplace}) is defined as the semisum of the integrals along two lines parallel to the real positive axis $u\ge 0$, slightly above and below it,  is consistent with some of the analytic properties of the true function $\wh D(s)$, in particular the absence of cuts on the spacelike axis $s<0$ ouside the Landau region. This prescription was adopted actually in  \cite{Costin1995,Costin_Duke, Costin_Book} for the Borel summation of the perturbative part, and also in the previous studies of the method of conformal mapping, which we use in the present work.
 
The series (\ref{eq:B}) converges  in the disk $|u|<1$, limited by the first ultraviolet renormalon at $u=-1$. On the other hand, the Laplace-Borel integral (\ref{eq:Laplace}) includes the range  $u>1$, where the series (\ref{eq:B}) is divergent. This is the  reason of the divergence of the original series (\ref{eq:hatD1}), obtained formally by inserting  (\ref{eq:B}) in (\ref{eq:Laplace}) and integrating term by term.

The domain of convergence  can be enlarged  by reexpanding  the function $B_{\wh D}(u)$ in powers of the variable which achieves the conformal mapping of the original complex $u$ plane onto the unit disk of a new complex plane. 
 This mapping, written for the first time in \cite{Caprini:1998wg},  has the form
\be\label{eq:w}
\tilde w(u)=\frac{\sqrt{1+u}-\sqrt{1-u/2}}{\sqrt{1+u}+\sqrt{1-u/2}}.
\ee
One can check that  $\tilde w(u)$ maps the complex  $u$ plane cut along the real axis for $u\ge 2$ and $u\le -1$ onto the interior of the circle $\vert w\vert\, =\, 1$ in the complex plane $w\equiv \tilde w(u)$,  such that  the origin $u=0$ of the $u$ plane
corresponds to the origin $w=0$ of the $w$ plane, and the upper (lower) edges of the cuts are mapped onto the upper
(lower) semicircles in the  $w$ plane.  

Consider now the expansion of $B_{\widehat D}(u)$ in powers of the variable $w$:
\be\label{eq:Bw} B_{\widehat D}(u)=\sum_{n=0}^\infty c_n \, w^n, \quad\quad w \equiv \tilde w(u),\ee
 where the coefficients $c_{n}$ are obtained from  $b_{k}$,
$k\leq n$, using Eqs. (\ref{eq:B}) and  (\ref{eq:w}). We recall that these  coefficients  are real, since $b_k$ are real and the function $\tilde w(u)$ defined in (\ref{eq:w}) is real analytic. For completeness, we quote the values of the low-order coefficients $c_n$, obtained using the perturbative coefficients $c_{n,1}$ given in (\ref{eq:cn1}):
\be\label{eq:ccoef} 
c_0=1,\quad c_1=1.94,\quad c_2=5.77, \quad c_3=18.50.
\ee

 We emphasize that, by expanding $B_{\widehat D}(u)$ according to (\ref{eq:Bw}), one makes full use of its
holomorphy domain, because the known part of it
(the first Riemann sheet) is
mapped onto the convergence  disk.  Therefore, the series (\ref{eq:Bw}) converges in the whole $u$ plane, up to the cuts along the real axis, i.e., in a much larger domain than the original series (\ref{eq:B}).  

By inserting the  expansion (\ref{eq:Bw})   of the Borel transform  in the Laplace-Borel integral (\ref{eq:Laplace}) regularized by PV prescription, and permutting the order of the integration and summation, we obtain new perturbative series for the Adler function in the complex $s$ plane. For convenience, we use  the notation from \cite{Caprini:2000js, Caprini:2023tfa},  writing
\be\label{eq:DI} 
\wh D(s)= \frac{1}{\beta_0} I(\beta_0 a_s(-s)).
\ee
Here $I$ denotes the series
\be\label{eq:cI}
I(a)=\sum_{n=0}^\infty c_n I_n (a),
\ee
where the coefficients $c_n$ appear in (\ref{eq:Bw}) and the expansion functions are
\be\label{eq:In}
I_n(a)=\text{PV} \int_0^\infty (\tilde w(u))^n e^{-u/a} du.
\ee

As proved in \cite{Caprini:2000js, Caprini:2023tfa}, unlike the original series (\ref{eq:hatD1}) which is divergent, the expansion (\ref{eq:cI}) converges in the complex $s$ plane, in particular along the circle $|s|=m_\tau^2$, if some conditions are fulfilled. On the other hand, the expansion functions  $I_n(a)$  are analytic in the complex $a$ plane and bounded for $\mbox{Re}\, a>0$, but have a cut along the axis $a<0$ and an essential singularity  [$\sim \exp(-1/a)$] at the origin $a=0$. As a consequence, when expanded in powers of $a$, $I_n(a)$ have divergent expansions, with coefficients exhibiting factorial growth. 

  \section{Power corrections to the standard perturbative series}\label{sec:OPE}
The presence of power-suppressed corrections to the standard series  (\ref{eq:hatD1}) 
has been advocated for the first time in \cite{Shifman:1978bx} by extending the validity of OPE to the nonperturbative regime.  For the Adler function, these corrections are parametrized as
\be\label{eq:ope}
D^{\text{OPE}}(s)=\frac{C_{4,0}(\alpha_s(-s))}{s^2}\langle {\cal O}_{4,0}\rangle+\sum\limits_{d=6}^\infty\frac{1}{(-s)^{d/2}}\sum_i C_{d,i}(\alpha_s(-s)) \langle {\cal O}_{d,\gamma_i}\rangle.
\ee
Here the terms $\langle {\cal O}_{d,\gamma_i}\rangle $ are nonperturbative vacuum matrix elements of light quark and
gluon field operators with anomalous dimensions $\gamma_i$,  called condensates, and the
functions $C_{d,i}$ are the corresponding Wilson coefficients,  computed perturbatively in terms of the coupling $\alpha_s$.

For massless quarks, the leading   correction with dimension $d = 4$  consists of a single term, related to the well-known gluon condensate matrix element $\langle a_s G^2\rangle$. To leading order in $\alpha_s$ and setting $s=-Q^2$, the dominant term in  (\ref{eq:ope}) is usually parametrized  as \cite{Beneke:2008ad, Benitez-Rathgeb:2022yqb}
\be\label{eq:GC}
D^{\text{OPE}}_{4}(-Q^2)= \frac{2 \pi^2}{3} \,  \frac{\langle a_s G^2\rangle}{Q^4},
\ee 
where we recall the standard value  $\langle a_s G^2\rangle = 0.012 \, \text{GeV}^4$ \cite{Shifman:1978bx}.
 Factors depending logarithmically on $Q^2$ are produced in (\ref{eq:GC}) by higher perturbative corrections in the Wilson coefficients.

The mathematical approach to resurgence developed in \cite{Costin1995, Costin_Duke, Costin_Book} requires the knowlwdge of the Borel transform of the perturbative series. The perturbative part is expressed as the Prinvipal Value of the Laplace-Borel integral, and the nonperturbative corrections are obtained from the discontinuity of the integral across the positive semiaxis. Therefore,  the infrared renormalons predict the form of the power corrections, up to an overall unknown constant.  The application of the algorithm to full QCD is not possible, since the exact form of the Borel transform is not known.
To illustrate the method, let us assume that  $B_{\widehat D}(u)$ consists from a single infrared renormalon, taken to be a branch-point at $u=2$:
 \be\label{eq:branch}
 B_{\widehat D}(u)= \frac{N}{(2-u)^\gamma},
\ee
where $\gamma$ is a positive noninteger number and the residue $N$ is real. 

 The discontinuity of the Laplace-Borel integral is given by  \cite{Beneke:2008ad}
\be\label{eq:disc}
\frac{N}{2 i} \left[\int\limits_0^\infty  
\frac{e^{-u/(\beta_0 a_s)}} {2-u-i 0}\, d u-\int\limits_0^\infty  
\frac{e^{-u/(\beta_0 a_s)}}{2-u+i 0}\, d u\right]=\frac{N}{\Gamma(\gamma)} (\beta_0 a_s)^{1-\gamma} e^{-2/(\beta_0 a_s)}.
\ee
According to the algorithm presented in \cite{Costin1995, Costin_Duke, Costin_Book},  the term generated by resurgence, which supplements the perturbative part, is equal to this discontinuity multiplied by an arbitrary real constant 
 $C$. Using for simplicity the one-loop solution of the renormalization-group equation (\ref{eq:rge}), which we parametrize  in terms of the QCD parameter $\Lambda$ as
\be\label{eq:oneloop}
a_s(Q^2)=\frac{1}{\beta_0 \ln\frac{Q^2}{\Lambda^2}}.
\ee
we obtain from (\ref{eq:disc})
\be\label{eq:resurg}
D^\text{resurg}_4(-Q^2)=\frac{NC}{\Gamma(\gamma)}\ (\ln{Q^2/\Lambda^2})^{1-\gamma}\,\frac{\Lambda^4}{Q^4},
\ee
where the subscript indicates the dimension of the power correction.

This expression, derived in a formal way, contains an unknown real constant $C$. We combine this constant with the normalization $N$ and  the QCD parameter $\Lambda$, to define  a parameter $G^2$ by
\be\label{eq:defG}
\frac{N C}{\Gamma(\gamma)}\,\Lambda^4= G^2.
\ee
With this definition,  the expression (\ref{eq:resurg}), predicted formally by resurgence, becomes
\be\label{eq:resurgG}
D^\text{resurg}_4(-Q^2)= (\ln{Q^2/\Lambda^2})^{1-\gamma}\,\frac{G^2}{Q^4}.
\ee

Compared to the standard power correction (\ref{eq:GC}) predicted by OPE, one may note the additional logarithmically-dependent factor present to leading order in (\ref{eq:resurgG}). As for the dimension-4 constant $G^2$, it is an arbitrary parameter which cannot be interpreted {\em a priori} as a gluon condensate. This shows that the power corrections depend on the form in which the perturbative part is written: for the standard truncated  series in powers of the coupling, the power corrections are assumed to have the expression (\ref{eq:GC}),  while  for a single infrared renormalon of a generic branch-point form,  the power correction is given by (\ref{eq:resurgG}). One may expect a different form of the power corrections for the modified perturbative series based on conformal mapping of the Borel plane. We shall investigate this problem in the next section.

\section{Power corrections to the expansion based on conformal mapping}\label{sec:OPE1}
We consider now the perturbative expansion based on conformal mapping of the Borel plane, defined in Eqs. (\ref{eq:DI}), (\ref{eq:cI}) and (\ref{eq:In}). As proved in \cite{Caprini:2023tfa}, the series  (\ref{eq:cI}) is convergent when some conditions are fulfilled. In particular, if the Borel transform $B_{\wh D}$ consists from a finite number of poles or branch points on the real axis of the Borel plane, the series converges along the circle $|s|=m_\tau^2$.  On the other hand,  the expansion functions  $I_n(a)$  are no longer powers of the coupling, as in the standard perturbation theory, but complicated functions singular at $a=0$, defined by means of the Laplace-Borel integral (\ref{eq:In}). As mentioned at the end of Sec. \ref{sec:Adler}, each function $I_n(a)$, when expanded in powers of $a$,  has a divergent expansion, with coefficients exhibiting factorial growth. Therefore, although the expansion (\ref{eq:cI}) is convergent, resurgence theory predicts the existence of additional terms to it. As in the previous section, for obtaining these terms we apply the algorithm from \cite{Costin1995, Costin_Duke, Costin_Book}. The application is straightforward, since the perturbative series (\ref{eq:cI}) is expressed as an exact Laplace-Borel integral, as required in  \cite{Costin1995, Costin_Duke, Costin_Book}.

We write the perturbative expansion of the Adler function  for $s=-Q^2$ as
\be\label{eq:DN}
\widehat D_\text{conf}(-Q^2)= \sum\limits_{n=0}^\infty c_n \, \frac{1}{\beta_0} \text{PV} \int_0^\infty  e^{-u/(\beta_0 a_s(Q^2))} (\tilde w(u))^n du. 
\ee
The function $\tilde w(u)$, defined in (\ref{eq:w}), has a branch point at $u=2$, and becomes complex for $u>2$. Moreover, since  the region $u\ge 2$ is mapped onto  the unit circle in the $w$ plane, we can write, for $u\ge 2$,
\be\label{eq:wpsi}
\tilde w(u)=e^{\pm i \psi(u)},
\ee
where the $\pm$ signs correspond to the upper (lower) edge of the cut and
\be\label{eq:psi}
\psi(u)=\arctan \frac{2\sqrt{1+u} \sqrt{u-2}}{3}. 
\ee
The discontinuity (the imaginary part) of the integrand in (\ref{eq:DN}) is obtained easily by noting that, for $u\ge 2$,
\be\label{eq:imwn}
\text{Im} (\tilde w(u))^n=\sin n\psi(u).
\ee

These expressions follow from the fact that the infrared renormalons are mapped on the unit circle of the $w$ plane. We note that the ultraviolet renormalons are also mapped on the unit circle, but their contribution is real for $u\ge 2$.  This is seen from Eq. (\ref{eq:w}), where the relevant factor $\sqrt{1+u}$ 
 is real for $u\ge -1$. On the other hand,  this factor enters  the expression of the
 discontinuity, as seen from (\ref{eq:psi}) and (\ref{eq:imwn}). So, the ultraviolet renormalons 
show their presence in the expression of the power corrections.

According to the algorithm from \cite{Costin1995, Costin_Duke, Costin_Book}, in order to obtain the term to be added to the perturbative part, we must calculate  the Laplace-Borel integral of the discontinuity and multiply the result by a constant (real in the present case). The integrand has only one singularity on the positive axis, at $u=2$. Therefore, we can write
\be\label{eq:resurg0}
D_\text{conf}^\text{resurg}(-Q^2)=\sum\limits_{n=0}^\infty c_n\frac{C'}{\beta_0} \int_2^\infty  e^{-u/(\beta_0 a_s(Q^2))} 
\sin n\psi(u)\, du, 
\ee
where $C'$ is an arbitrary real constant.
By making the change of variable
\be
t=u-2
\ee
we write (\ref{eq:resurg0}) as
\be\label{eq:resurgconf}
D_\text{conf}^\text{resurg}(-Q^2)=\frac{C'}{\beta_0}  e^{-2/(\beta_0 a_s(Q^2))} \sum\limits_{n= 0}^\infty c_n\int_0^\infty  e^{-t/(\beta_0 a_s(Q^2))}  \sin n\psi(t) \,dt, 
\ee
where
\be\label{eq:psit}
\psi(t)=\arctan \frac{2\sqrt{3+t} \sqrt{t}}{3}.
\ee

As in the previous section,  in the one-loop approximation  (\ref{eq:oneloop}) of the coupling, the exponential in front of the integral can be written as $\Lambda^4/Q^4$, leading to
\be\label{eq:resurg1}
D_\text{conf}^\text{resurg}(-Q^2)= \frac{\Lambda^4}{Q^4}\frac{C'}{\beta_0} \, \sum\limits_{n= 0}^\infty c_n\,\int_0^\infty  e^{-t/(\beta_0 a_s(Q^2))} \sin n\psi(t) \,dt.
\ee 

The integrals appearing in this equation are nonzero for $n\ge 1$ and depend logarithmically on $Q^2$. We can obtain an analytic expression of these integrals, valid with good approximation,  by noting that,  
for small coupling $a_s$, the contribution of large $t$ to the integrals is suppressed. Therefore, we can use the approximate expression
\be
\psi(t)\approx \frac{2}{\sqrt{3}}\, \sqrt{t},
\ee
which follows from (\ref{eq:psit}) at small $t$. By inserting
\be
\sin n\psi(t)= \frac{1}{2i}\,\left[e^{i \frac{2 n}{\sqrt{3}}\,\sqrt{t}}-e^{-i \frac{2 n}{\sqrt{3}}\,\sqrt{t}}\right]
\ee
in the r.h.s. of (\ref{eq:resurg1}), we obtain the exact result
\be
\int_0^\infty  e^{-t/(\beta_0 a_s(Q^2))} \sin n\psi(t) \,dt =n \sqrt\frac{\pi}{3} \,(\beta_0 a_s(Q^2))^{3/2} \,e^{-\frac{n^2}{3} \beta_0 a_s(Q^2)}.
\ee
Then (\ref{eq:resurg1}) can be written as
\be\label{eq:Dresurg}
D_\text{conf}^\text{resurg}(-Q^2)= C' \frac{\Lambda^4}{Q^4}  \sqrt\frac{\pi \beta_0}{3} \, F(Q^2),
\ee
where
\be\label{eq:F}
F(Q^2)=(a_s(Q^2))^{3/2}\,\sum_{n= 1}^\infty n\, c_n\, e^{-\frac{n^2}{3} \beta_0 a_s(Q^2)}.
\ee

The power correction (\ref{eq:Dresurg}), derived in a formal way,  contains the unknown constant $C'$. As in the previous section,   we define a dimensionfull parameter $G^2_\text{conf}$  by the relation
\be\label{eq:defGconf}
C' \Lambda^4 \sqrt\frac{\pi \beta_0}{3} =  G^2_\text{conf} ,\ee
similar to (\ref{eq:defG}). Then the dimension $d=4$ power correction in the scheme based  on conformal mapping of the Borel plane, predicted by resurgence, is written as
\be\label{eq:OPEconf}
D_\text{4, conf}^\text{OPE}(-Q^2)= \frac{F(Q^2)}{Q^4}\, G^2_\text{conf}.
\ee
Compared to the standard power correction (\ref{eq:GC}), the expression (\ref{eq:OPEconf}) contains the additional function $F(Q^2)$,  defined in (\ref{eq:F}). Also, the arbitrary parameter $G^2_\text{conf}$ appearing in this expression cannot be interpreted as a gluon condensate. 
As shown already in the previous section,  the expression of the power corrections depend on the form in which the perturbative part is written.  

Before ending this section, it is of interest  to investigate the convergence of the series appearing in the definition (\ref{eq:F}) of the function $F$. As discussed in \cite{Caprini:2023tfa}, if the Borel transform $B_{\wh D}(u)$ 
consists from a finite sum of poles or branch points on the real axis, the coefficients $c_n$ of the expansion (\ref{eq:Bw}) are bounded as $|c_n|<\exp(\sqrt{n})$, and the perturbative series (\ref{eq:cI}) is convergent along the circle $|s|=m_\tau^2$. In order to test the convergence of the series (\ref{eq:F}) along the circle,  we consider the ratio
\be\label{eq:ratio}
r_n=\frac{ n c_n \exp(-n^2 \xi)}{(n-1) c_{n-1}\exp(-(n-1)^2 \xi)},
\ee
where $\xi= \beta_0 \text{Re}\, a_s(-s)/3$. Using the above estimate of $c_n$ we obtain, for large $n$,
\be
r_n\sim \exp(1/(2\sqrt{n})-2 n\xi)
\ee
which implies
\be\label{eq:rlim}
\lim_{n\to\infty} r_n<1,
\ee 
since $\xi>0$ for $s$ on the circle $|s|=m_\tau^2$. According to the ratio test, the series appearing in (\ref{eq:F}) is convergent. Therefore, for a finite number of poles and branch points in the Borel transform $B_{\wh D}$, both the modified perturbative series (\ref{eq:cI}) and the series entering the $d=4$  power correction  (\ref{eq:F})  are convergent in a region of the $s$ complex plane, in particular on the circle $|s|=m_\tau^2$. We shall exemplify numerically this statement in the next section.

\section{Numerical and phenomenological implications}\label{sec:phen} 

\begin{figure}\vspace{0.2cm}
\includegraphics[width=7.5cm]{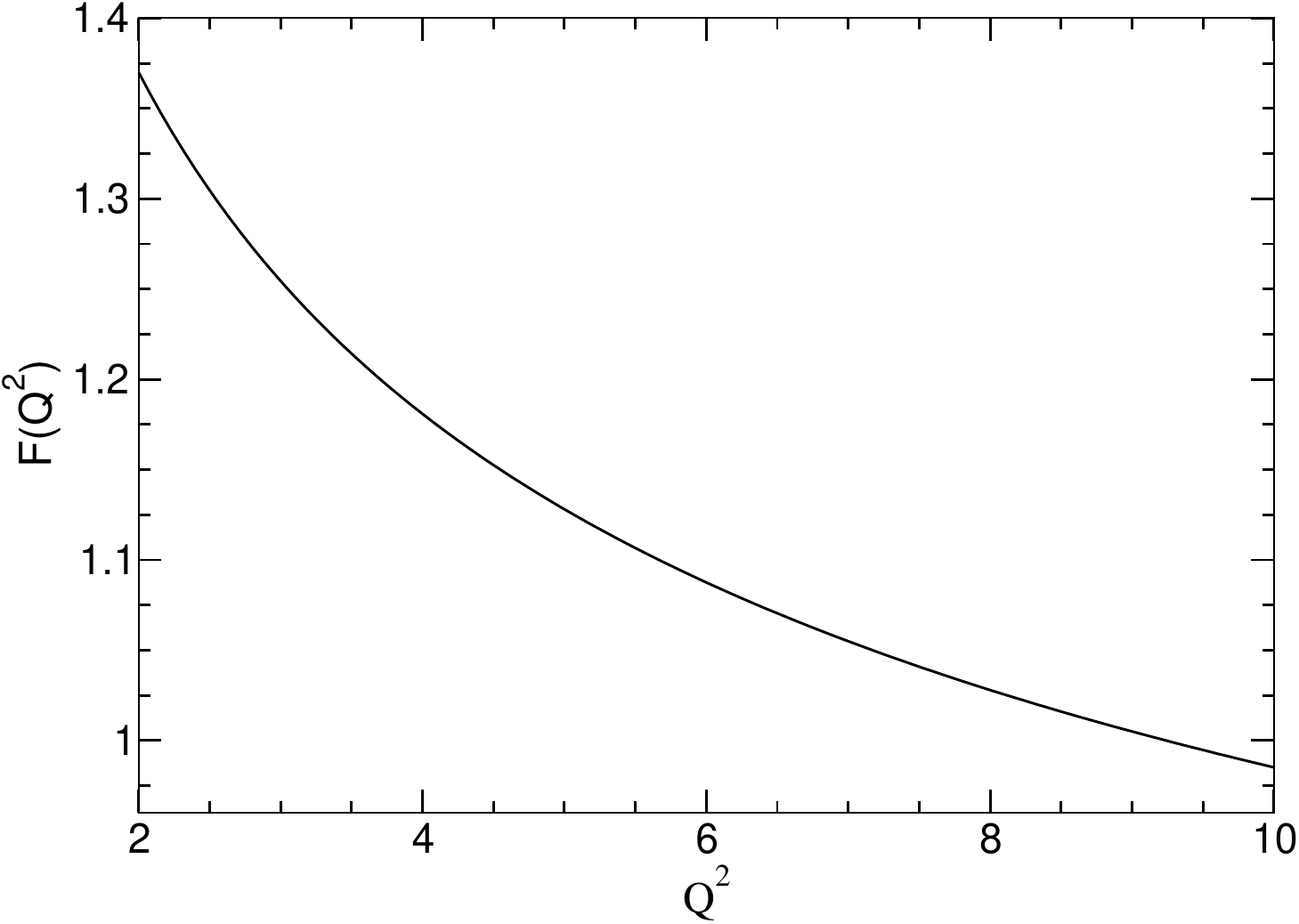}
\caption{The function $F(Q^2)$ defined in (\ref{eq:F}), truncated at $n\le 3$,  on  the Euclidian axis  $Q^2>0$.}
\label{fig:F}
\end{figure}

\begin{figure}\vspace{0.2cm}
\includegraphics[width=7.5cm]{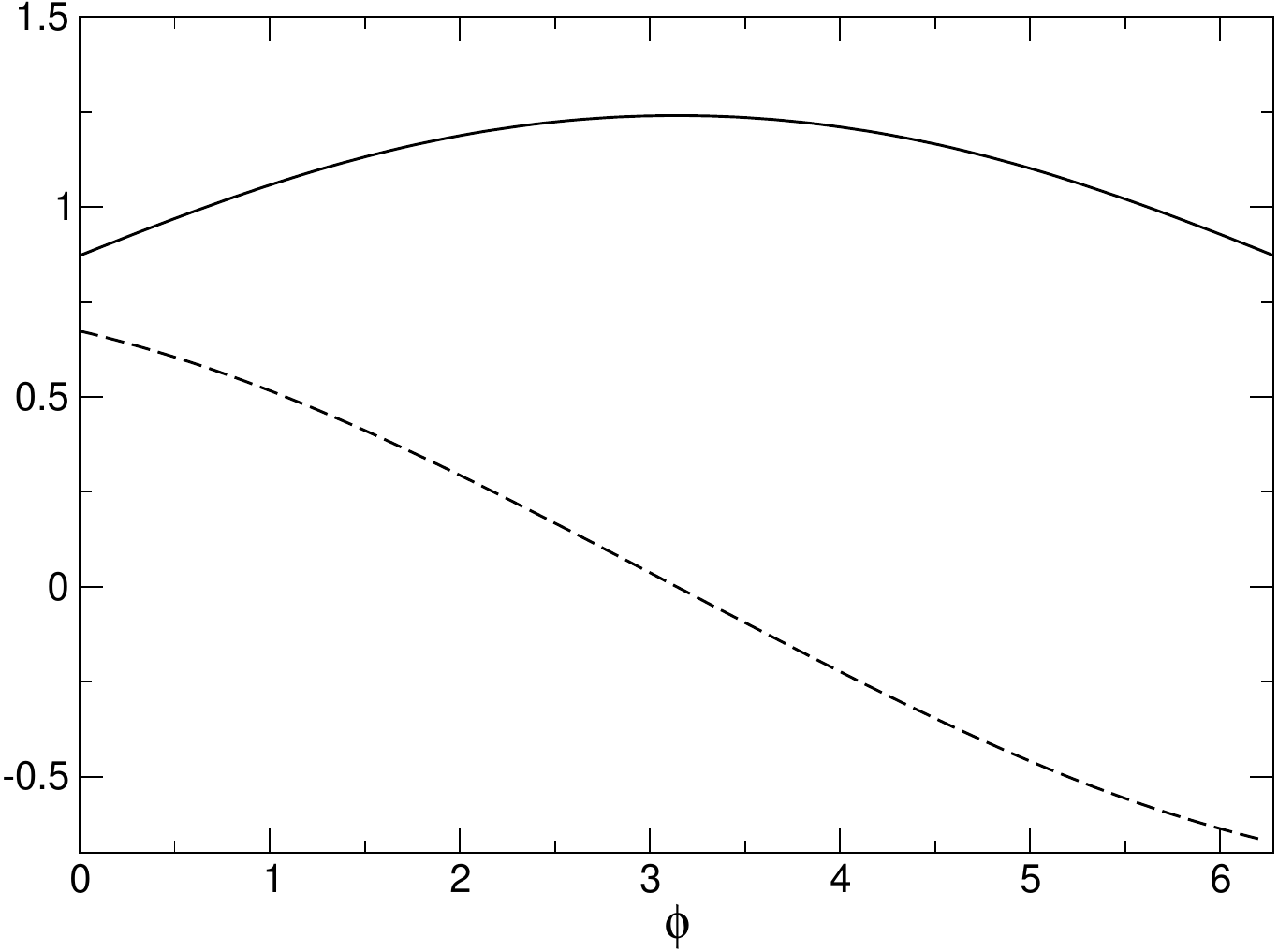}
\caption{Real part (solid line) and imaginary part  (dashed line) of the function $F(-s)$,  along the circle $s=m_\tau^2 e^{i\phi}$.}
\label{fig:Fcomp}
\end{figure}

To illustrate the properties of the function $F$,  we show first in Fig. \ref{fig:F} the expression (\ref{eq:F}) with the series truncated at $n\le 3$, calculated  using the low-order coefficients $c_n$ from (\ref{eq:ccoef}) and the one-loop coupling (\ref{eq:oneloop}) with $\Lambda=0.2\, \text{GeV}$, which gives $\alpha_s(m_\tau^2)=0.32$,  for $Q^2$ in a range of the Euclidian axis. One may note the slow decrease of $F(Q^2)$  with increasing $Q^2$.
In Fig. \ref{fig:Fcomp} we show the real and the imaginary parts of $F$, calculated with the same input,  for complex $Q^2=-s$ along the circle $s=m_\tau^2 e^{i\phi}$, as functions of $\phi$.

Our results have implications for the calculation of the spectral function moments
\be\label{eq:delta0W}
\delta^{(0)}_W=\frac{1}{2 \pi i}\,\oint\limits_{|s|=s_0} \frac{ds}{s} W(s/s_0)  \wh D(s),
\ee
which are important for the determination of the strong coupling $\alpha_s$ from the data  on hadronic $\tau$ decay.
Since $\oint \frac{ds}{s} \frac{s^j}{s^2}$ is nonzero only for $j=2$,  
the contribution of the standard power correction (\ref{eq:GC}) due to the gluon condensate vanishes identically if the weight function $W$ contains only terms $s^j$ with $j\ne 2$.  In  \cite{Benitez-Rathgeb:2022yqb}, these moments are referred to as  ''gluon condensate suppressed moments``. An example is the kinematical weight, $W_{\tau}(x)=(1-x)^3(1+x)$, entering the expression of the total hadronic decay rate of the $\tau$ lepton.

\begin{figure}\vspace{0.2cm}
\includegraphics[width=7.5cm]{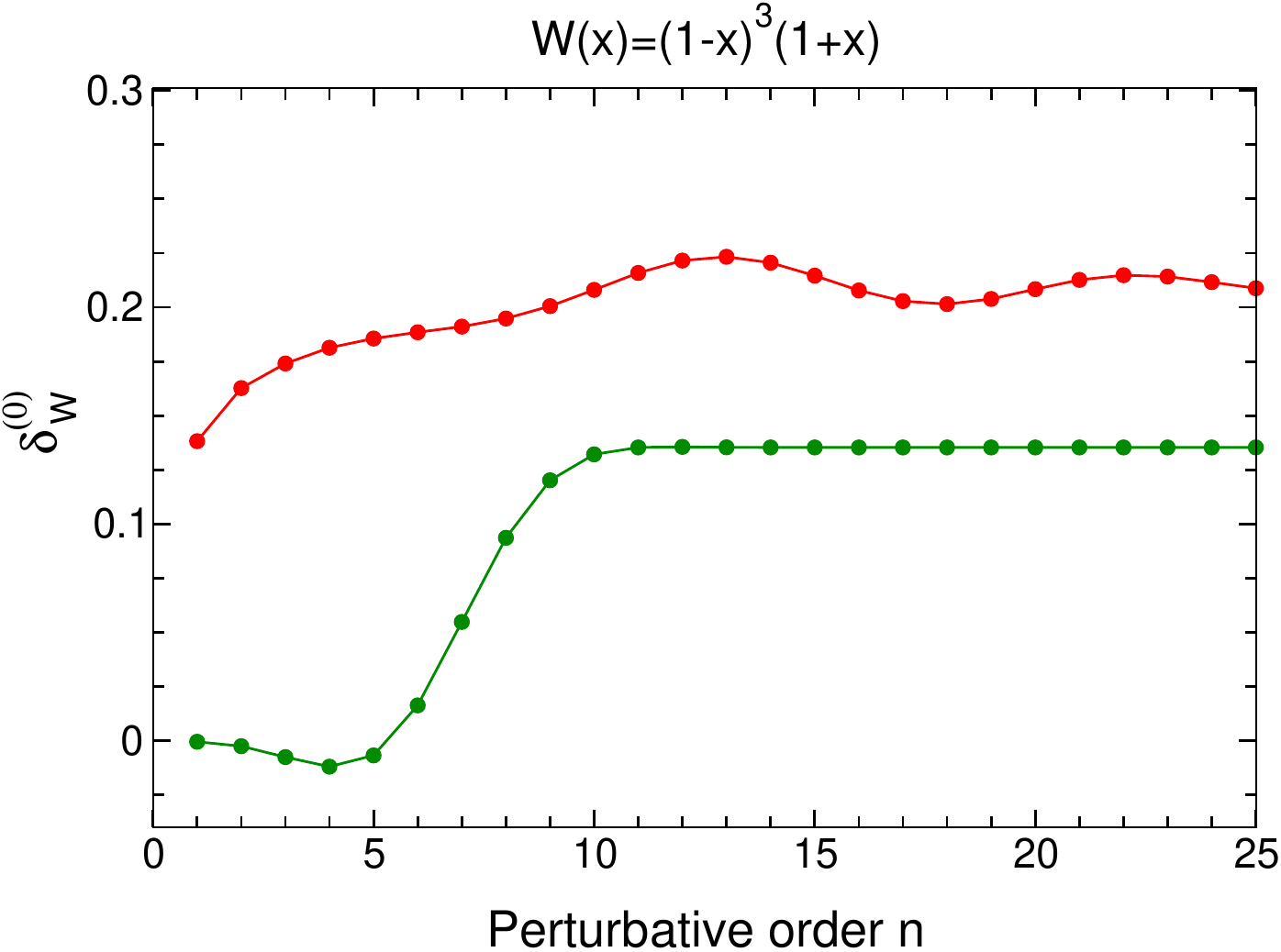} \hspace{0.5cm}\includegraphics[width=7.5cm]{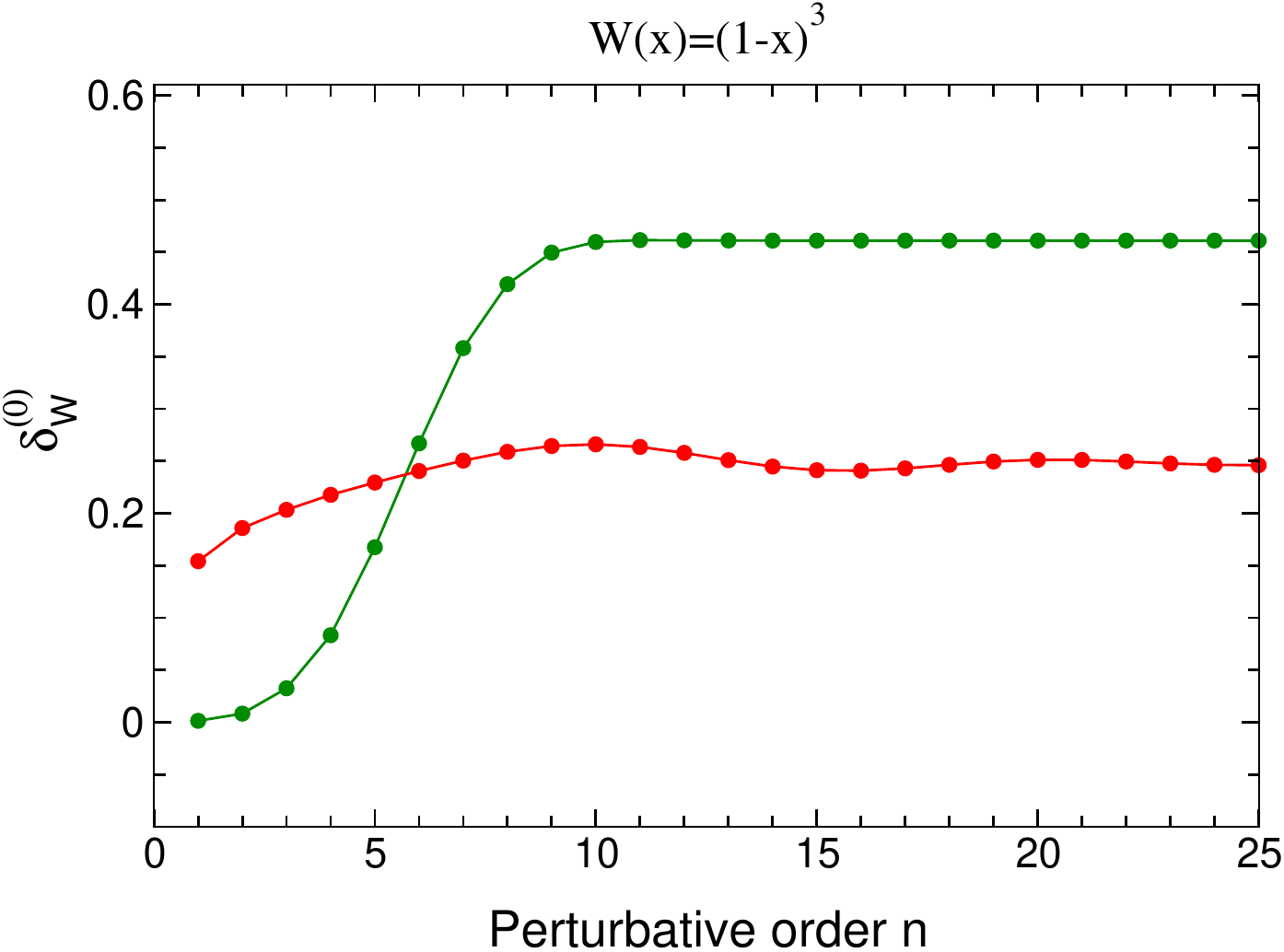}\vspace{0.7cm}
\includegraphics[width=7.5cm]{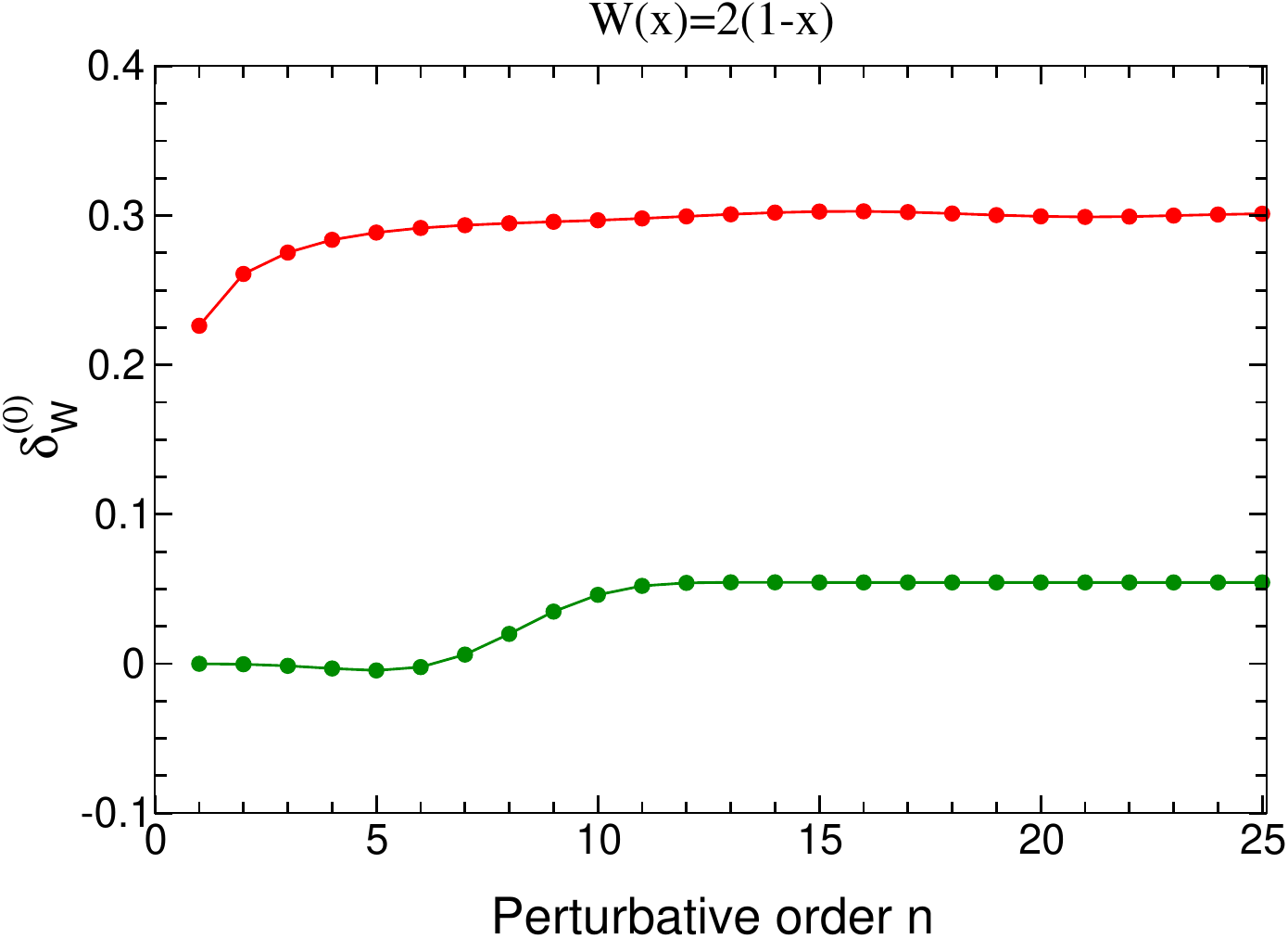} \hspace{0.5cm}\includegraphics[width=7.5cm]{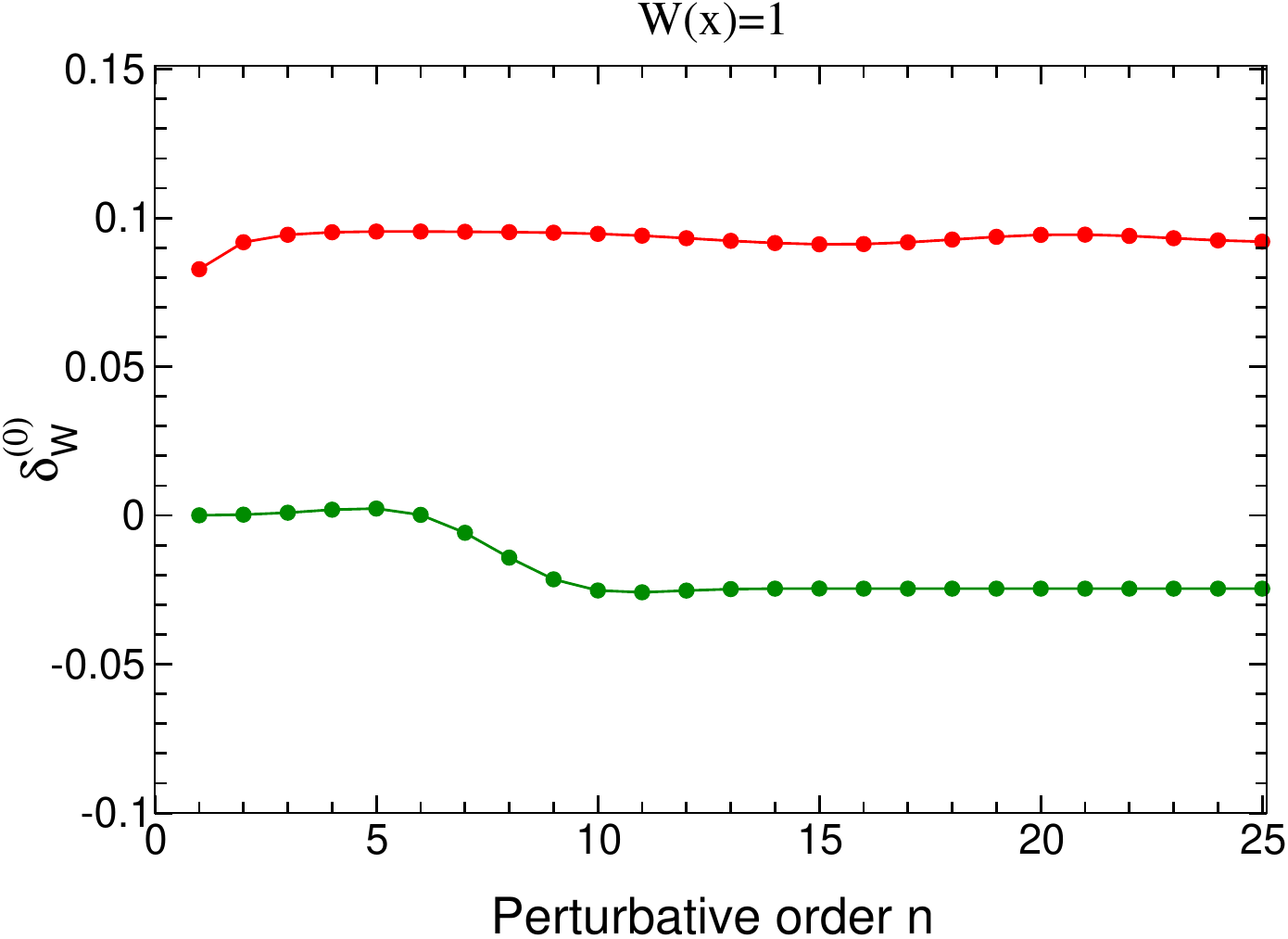}
\caption{Spectral function moments in the framework based on conformal mapping: perturbative series (red points) and  power corrections  (green points),  as functions of the perturbative order $n$. The multi-renormalon Borel model proposed in \cite{Beneke:2008ad} was used for generating the higher-order perturbative coefficients.}
\label{fig:delta0}
\end{figure}

The presence of logarithmic factors in the expression of the power correction  modifies this result,  leading to  nonzero values also for terms $s^j$ with $j\ne 2$.  For the standard OPE, these contributions are tiny  and can be neglected in practical applications \cite{Beneke:2008ad, Benitez-Rathgeb:2022yqb}. On the other hand, due to the function $F$, the contribution of the power correction (\ref{eq:OPEconf}) to moments with $j\ne 2$ can be more substantial. 

For illustration, we shall investigate the  spectral function moments in the framework based of conformal mappings,  using for generating the higher-order perturbative coefficients a  model proposed in \cite{Beneke:2008ad} and used afterwards in many studies of the QCD Adler function  (see for instance \cite{Benitez-Rathgeb:2022yqb, Benitez-Rathgeb:2022hfj, Beneke:2023wkq, Caprini:2009vf, Caprini:2011ya, Abbas:2013usa, Caprini:2020lff}). In this model, the Adler function is defined as a PV-regulated  Laplace-Borel integral, with  the Borel transform written in terms of the few infrared and ultraviolet renormalons 
\be\label{eq:BBJ}
B_D(u)=B_1^\text{UV}(u) +  B_2^\text{IR}(u) + B_3^\text{IR}(u) +d_0^\text{PO} + d_1^\text{PO} u,
\ee
parametrized as
\be\label{eq:BIR}
B_p^\text{IR}(u)= \frac{d_p^\text{IR}}{(p-u)^{\gamma_p}}\,
\left[\, 1 + \tilde b_{1,p} (p-u)  +\ldots \,\right],
\ee
\be\label{eq:BUV}
B_p^\text{UV}(u)=\frac{d_p^\text{UV}}{(p+u)^{\bar\gamma_p}}\,
\left[\, 1 + \bar b_{1,p} (p+u)  +\ldots \,\right].
\ee
Here  $\gamma_p, \bar\gamma_p, b_{1, p}$ and $\bar{b}_{1, p}$ are determined from renormalization-group arguments 
 and the free parameters have been obtained in  \cite{Beneke:2008ad} as 
\bea 
&& d_1^\text{UV}=-\,1.56\times 10^{-2},\,\,
d_2^\text{IR}=3.16,\,\,
d_3^\text{IR}=-13.5,\nn\\
&& d_0^\text{PO}=0.781, \,\,
d_1^\text{PO}=7.66\times 10^{-3},
\eea
by the requirement to reproduce the perturbative coefficients  $c_{n,1}$ in $\overline{\text{MS}}$ scheme for $n\le 4$, given in (\ref{eq:cn1}), and the estimate $c_{5,1}=283$.

In Fig. \ref{fig:delta0} we present the results of our analysis for  the kinematical moment and other three moments considered in recent studies. Except the second one, $W(x)=(1-x)^3$, all are  ''gluon condensate suppressed moments``, i.e., do not contain terms proportional to $s^2$. With red points we present the results of the perturbative series improved by conformal mappings, given above in Eqs. (\ref{eq:DI}), (\ref{eq:cI}) and (\ref{eq:In}). We  emphasize that this corresponds to the CIPT formulation, which, as discussed in \cite{Caprini:2023tfa}, has better properties than FOPT for expansions  based on conformal mappings.  With green points we show the contributions of the power correction  (\ref{eq:OPEconf}), with the function $F$ defined by the series  (\ref{eq:F}). In both cases the series have been truncated at the order $n$. We took  $s_0=m_\tau^2$ and used 
$\alpha_s(m_\tau^2)=0.32$. Finally, we adopted for the normalization of  (\ref{eq:OPEconf}) the value $G^2_\text{conf}=0.01\,\text{GeV}^4$. We recall that this parameter is arbitrary, and the magnitude similar to that of the standard gluon condensate quoted below (\ref{eq:GC}) has no particular significance.

The results show the impressive convergence of both the perturbative and nonperturbative series up to a high truncation order $n$. We recall that for the perturbative part the convergence was proved in \cite{Caprini:2023tfa} and was noticed numerically in the previous papers \cite{Caprini:2009vf, Caprini:2011ya, Abbas:2013usa, Caprini:2020lff}, while for the nonperturbative part the convergence was proved in the previous section of this paper.

We note also that, for the reasonable choice of $G^2_\text{conf}$ quoted above, the contribution of the power corrections is considerably smaller than that of the perturbative part. The only exception is the second moment, where the nonperturbative contribution becomes larger than the perturbative one above a certain order $n$.  This is explained by the fact that the weight of this moment contains a term proportional to $s^2$, i.e., it is not a  ''gluon condensate suppressed moment`` in the terminology of  \cite{Benitez-Rathgeb:2022yqb}. Thus, the pattern noticed for the contribution to the moments of the standard OPE is preserved qualitatively by the power correction (\ref{eq:OPEconf}), with of course some corrections brought by the function $F$. The remark is useful for choosing the adequate moments for a phenomenological analysis of the data on hadronic decay of the $\tau$ lepton, which will allow also the determination of the normalization $G^2_\text{conf}$.

\section{Summary and conclusions}\label{sec:conc} 
In the present work we investigated the power corrections to the modified perturbative series (\ref{eq:cI})   based on conformal mapping of the Borel plane. 
As recalled in Sec. \ref{sec:Adler}, the expansion functions defined in  (\ref{eq:In})  are singular at the origin of the coupling plane, exhibiting a nonperturbative behavior.  So, unlike the standard expansions in powers of the coupling, which are holomorphic functions of $a_s$  when truncated at finite orders, the modified perturbative expansion (\ref{eq:cI}) exhibits, even at finite orders, a singular behavior, much like  the expanded function itself.  Therefore, it is reasonable to expect that the additional  nonperturbative terms,  if present, will be different from the standard power corrections  (in \cite{Caprini:2020lff} it was even assumed that these terms can be neglected). 

The power corrections to the standard perturbative series in QCD, proposed for the first time in \cite{Shifman:1978bx}, 
have been obtained by extending the operator product expansion to the nonperturbative regime. 
For the expansions based on conformal mapping, a physical argument for generating the power corrections is not evident. Therefore, we resorted to the mathematical resurgence theory, specifically to the algorithm described  in \cite{Costin1995, Costin_Duke, Costin_Book} for the solutions of differential equations. The algorithm requires the knowledge of the Borel transform of the  perturbative series, and expresses the additional nonperturbative contribution in terms of the discontinuity of this function across the positive axis of the Borel plane. The algorithm has been used in  \cite{Maiezza:2021mry, Caprini:2023kpw} for deriving the resurgent representation of the Adler function in the large-$\beta_0$ approximation \cite{Beneke:1992ch, Broadhurst:1992si, Beneke:1994qe}, when the exact expression of the Borel transform is known.

 In full QCD, the Borel transform of the standard perturbative expansion  in not known exactly. In Sec. \ref{sec:OPE} we illustrated the algorithm given in  \cite{Costin1995, Costin_Duke, Costin_Book} in the simple case of a single branch point at $u=2$,  showing that the nonperturbative terms depend on the form of the perturbative part.

In Sec. \ref{sec:OPE1} we derived the nonperturbative corrections to  the perturbative expansion (\ref{eq:cI})  based on conformal mapping of the Borel plane. In this case the perturbative expansion is already expressed as a Laplace-Borel integral, allowing the straightforward application of the algorithm of \cite{Costin1995, Costin_Duke, Costin_Book}.   The expression of the dominant power correction is given in Eq. (\ref{eq:OPEconf}).  Compared to the standard term (\ref{eq:GC}),  it contains the additional function $F(Q^2)$,  defined in (\ref{eq:F}) as a series, which was shown to converge. We emphasize also that  the quantity denoted as $G^2_\text{conf}$, defined in (\ref{eq:defGconf}),  is an arbitrary parameter, with no obvious physical interpretation. One can speculate that it should be small, since the perturbative expansion (\ref{eq:cI}) already captures the singular behavior of the Adler function at $a_s=0$.

The properties of the nonperturbative corrections derived in this paper have been discussed in Sec. \ref{sec:phen} for several spectral moments of interest for the determination of the strong coupling from hadronic $\tau$ decays. Using a Borel model for generating higher-order perturbative coefficients of the Adler function, we obtained a remarkable convergence of both the perturbative series and the nonperturbative one in the framework based on conformal mapping. Also, for a reasonable normalization of the power corrections, their contribution is much smaller than the perturbative one for all the  moments where also the standard gluon condensate is suppressed.

The results of this paper, besides their conceptual interest, are relevant for the determination of the strong coupling from data on hadronic $\tau$ decay in the framework based on the conformal mapping of the Borel plane. We plan to present a detailed phenomenological analysis in a future work.


\end{document}